\documentclass{article}

\usepackage{graphicx,type1cm,eso-pic,color}

\makeatletter \AddToShipoutPicture{%
\setlength{\@tempdimb}{.5\paperwidth}%
\setlength{\@tempdimc}{.5\paperheight}%
\setlength{\unitlength}{1pt}%
\put(\strip@pt\@tempdimb,\strip@pt\@tempdimc){%

\makebox(-500,-0){\rotatebox{90}{\textcolor[gray]{0.92}%
{\fontsize{0.7cm}{0.7cm}\selectfont{\textcopyright
Copyright 2011 - John Scoville}}}} }%
} \makeatother

\begin{document}

\title{On Macroscopic Complexity and Perceptual
Coding}

\author{John Scoville}

\maketitle \begin{abstract} The theoretical limits of
'lossy' data compression algorithms are considered.
The complexity of an object as seen by a macroscopic
observer is the size of the perceptual code which
discards all information that can be lost without
altering the perception of the specified observer. The
complexity of this macroscopically observed state is
the simplest description of any microstate comprising
that macrostate. Inference and pattern recognition
based on macrostate rather than microstate
complexities will take advantage of the complexity of
the macroscopic observer to ignore irrelevant noise.
\end{abstract}

\section*{The quantification of information}
Information theory in its modern form originated from
Claude Shannon's\cite{SH48} usage of Gibbs' entropy
formula to describe communication channels:
\begin{equation} S = -k \sum{ P_{i} \log{P_{i}} }
\end{equation}

This formula originally applied to an ensemble of
microscopic states\cite{GI1902}, and the analytic form
of expected log-probability describes the entropy of
systems and their representations whether classical or
quantum in nature.  In the context of quantum
mechanics, it becomes the von Neumann entropy of the
state density matrix, $S = - trace (p \log{p})$. The
story goes that it was actually von Neumann who
suggested the term 'entropy' to Shannon for his
information function, for two reasons: 'In the first
place your uncertainty function has been used in
statistical mechanics under that name, so it already
has a name. In the second place, and more important,
nobody knows what entropy really is, so in a debate
you will always have the advantage.'

Entropy has the units of the logarithm of action\cite{LL80.1}
Shannon showed that, in the absence of Boltzmann's constant, $k$, entropy
quantifies the number of bits of data needed to
identify a sample from some distribution.  It
quantifies the amount of choice or uncertainty that
must be overcome in order to invoke the axiom of
choice and select a specific element from a set.  The
Shannon Entropy, $H$, limits the information capacity
of a signal communicated using an alphabet or codebook
with known distribution, P. \begin{equation} H = -
\sum{ P_{i} \log{P_{i}} } \end{equation}

Or, in the case of a continuous pdf, $H = - \int {
p(x) \log{p(x)} dx} $. The functional form of
information entropy has the properties that one
expects from a linear measure of choice\cite{SH48}
and, as such, establishes a theoretical bound for the
average information capacity of a string of symbols
sent from an ergodic source to a receiver.  The base
of the logarithm is equal to the cardinality of the
symbol set.  When the base is two, the units of
entropy are bits, in base $e$ they are 'nats', etc.

Shannon entropy represents choice or uncertainty in
the space of possible states; it is synonymous with
the development of information theory\cite{CO91}.
However, the definition of entropy due to Boltzmann is
still more widely known for its role in the
thermodynamics of physical
systems\cite{GI1902,RE65,KI80,FE56,PA96,PL45,LL80.1}
than its role in the theory of
information\cite{GA94,ZU89,LV97}.  The definition of
microscopic entropy (originally referred to by
Boltzmann as 'molecular chaos') is \begin{equation} S
= k \log{n} \end{equation}

For a set having n discrete elements.  This definition
may be extended to continuous spaces by considering n
as a volume V of phase space, having some measure,
$\mu$, in such a case the entropy becomes $S = k
\log{\mu(V)}$.

Boltzmann's thermodynamic entropy and Gibbs's
statistical mechanical entropy are closely related.
The Gibbs entropy of an ensemble reduces to the
Boltzmann entropy in the case of an ensemble of n
equally likely microscopic states (or a volume n of
state space) corresponding to a single maximum-entropy
equilibrium state.  Alternately, Gibbs' entropy
function may be shown to emerge from Boltzmann's
entropy as the increase in phase space volume
corresponds to the expected value of the
log-probability.\cite{LL80.1}

The Boltzmann entropy of a non-equilibrium system
(which could have any number of partial equilibrium
macrostates) is proportional to the sum of the
entropies of each macrostate.\cite{LL80.1}  The
Boltzmann entropy of such a macrostate is the
logarithm of the number of microscopic states
consistent with the observed macroscopic state of the
ensemble, or, equivalently, the volume of phase space
occupied by that macroscopic state.

Rather than Boltzmann's entropy function, which
includes a constant to describe the phase spaces of
physical systems, we will refer to an abstract
Boltzmann entropy function on a discrete state space,
$S = \log{n}$ which is simply the logarithm of the set
cardinality n.  If the set is continuous the form $S =
\log{\mu(V)}$ is implied.  A discrete version of this
abstract Boltzmann entropy takes the form of the
Hartley information function $\log{p}$. \cite{HA28}

Shannon entropy is defined for ergodic
sources\cite{SH48}, not particular instances of data.
If a source is not ergodic then the results of
classical information theory may hold only
approximately, on average, or
asymptotically.\cite{LV97}.  Colloquially, this means
that a series is statistically uniform throughout,
which is related to the notion of a stationary
stochastic process.  For example, for a memoryless
source such as a coin, $01101001111010101101$ may seem
like a typical sequence, but $11111111111100000000$
seems, offhand, highly unlikely to have been produced
by the same random process.  However, these sequences
have the same number of 1s and 0s, so when viewed as
an unordered set (rather than, for instance, a Markov
chain) they have identical distributions with the same
Shannon entropy.

Kolmogorov Complexity resolves this difficulty by
describing any type of binary symbolic information,
regardless of the source.  It is defined as the
minimum amount of information needed to completely
reconstruct some object, represented as a binary
string of symbols, X\cite{CO91,LV97}. \begin{equation}
C_{f}(X) = \min_{f(p)=X}|p| \end{equation}

In the parlance of computer science, f is a computer
and p is a program running on that computer.  The
Kolmogorov Complexity is the length of the shortest
computer program which terminates with X as output. In
the example of the binary sequences above, the second
clearly has a simpler algorithmic representation,
whereas the first is nearly random.

The Turing equivalence of different
computers\cite{TU36} relates Kolmogorov complexities
by a equivalence constant\cite{LV97,CO91}:
\begin{equation} C_{f}(X) = C_{g}(X) + C
\end{equation}

Since optimal specification does not depend on the
particular computer used, we will assume a standard
computer f unless otherwise specified.

The properties of $C(X)$ are sometimes more natural
when the set of possible X are constrained to be
prefix-free, that is, no X is a prefix of another X,
so programs are self-delimiting rather than being
demarcated by stop symbols.  In this case, we refer to
Chaitin's algorithmic prefix complexity $K(X)$.  We
won't delve into the details of $K(X)$, but we note
that a program can be made self-delimiting by
recursively prefixing the value of its length, and the
length of this prefix, and so forth, so $K(X) = C(X) +
C(C(X) + O(C(C(C(X))))$\cite{LV97}.  $K(X)$ gains some
important attributes\cite{LV97,CH87} that $C(X)$
lacks.  One is convergence of the universal
probability: \begin{equation} U(x) =
\sum_{f(p)=X}{2^{-|p|}} \end{equation}

This probability measure may be interpreted as the
probability that a randomly selected prefix-free
program terminates with x as its output. Convergence
is assured by the use of a self-delimiting prefix
code, as the Kraft inequality\cite{CO91} states that
the lengths of codewords x in a prefix code satisfy
$\sum 2^{l(x)} \leq 1$.  Though convergence to the
limit is in general very slow\cite{LV97}, this series
is often dominated by the shortest program, and $U(x)
\approx 2^{-K(x)}$ constitutes a reasonable
first-order approximation\cite{CO91,LV97} to the
universal probability for these typical objects which
are said to be 'shallow', whereas certain special
'deep' objects converge more slowly.

Performing complexity-based inference requires that we
have a notion of the information that one object
contains about another, or vice versa.  In the theory
of Kolmogorov complexity\cite{CH87,LV97}, this is
achieved by the conditional prefix complexity:
\begin{equation} K(A|B)=K(AB)-K(B) \end{equation}

Where the combination AB is the concatenation of
strings A and B. Algorithmic prefix complexity
characterizes measurements at finite precision and may
also be defined via algorithmic entropy, which we will
consider shortly.

The notion of splitting complexity into a 'regular'
(low prior probability) part and a 'random' (high
prior probability) part is also originally due to
Kolmogorov.  The Kolmologorov Minimal Sufficient
Statistic identifies the smallest superset of x which
may be described with less than k bits.  This is
closely related to the notion of stochastic processes
used earlier by Langevin to separate dynamical systems
into deterministic and random components.

The Kolmogorov complexity may be used to define
stochastic sequences in general, as such, is
fundamental to the notion of a statistical
probability. \cite{LV97}  For natural numbers $k$ and
$\delta$, we say that a string x is
$(k,\delta)$-stochastic if and only if there exists a
finite set $A$ such that: \begin{equation} x \in A,
C(A) \leq k, C(x|A)\geq log |A| - \delta
\end{equation}

The deviation from randomness, $\delta$, indicates
whether x is a typical or atypical member of A.  The
Kolmologorov Minimal Sufficient Statistic for x, given
$n=|x|$, is the set of minimum cardinality subject to
the first two constraints of stochasticity.  This is
defined through the Kolmologorov Structure Function,
$C_{k}(x|n)$: \begin{equation} C_{k}(x|n) = \min {
\{\log{|A|}: x \in A,C(A|n) \leq k\}} \end{equation}

The minimal set $A_{0}$ minimizes the randomness
deficiency, $\delta$, and is referred to as the
Kolmologorov Minimal Sufficient Statistic for x given
n.  This generalizes the notion of fitting a
distribution to x.  The Kolmogorov Structure Function
$C_{k}(x|n)$ measures the amount of randomness in the
string x.  For n coin tosses, it is nearly n, for a
number such as $\pi$, it is O(1).

Another fundamental partitioning of random and
nonrandom data is provided by the Algorithmic Entropy
function\cite{ZU89}, introduced by Zurek as physical
entropy as it generalizes classical
thermodynamics to a physical theory of information.
Algorithmic Entropy\cite{ZU89,LV97} combines
Kolmogorov complexity and Boltzmann entropy to measure
the macroscopic complexity of certain types of
measurements.  It relates computation and the
informatic content of real-valued measurements to
statistical mechanics.  The algorithmic entropy of a
string, H(Z) (not to be confused with the Shannon
Information, H, of a source) is defined in its most
basic form as: \begin{equation} H(Z) = K(Z) + S
\end{equation}

In this context, $Z=X_{1:n}$ is a description of a
macroscopic observation constructed by truncating a
microscopic state X to a bit string of length n. K(X)
is the algorithmic prefix complexity\cite{CH87,LV97}
of this representation of the macrostate.  In the case
of algorithmic entropy, the Boltzmann entropy S is
seen to be the additional complexity needed to specify
a microstate given knowledge of its macrostate.

Since all the microstates comprising a partition of
macrostate share a common prefix in their string
representation, K(X) is constructed as the prefix
complexity of these microstates.  The microstates are
contained in a volume of state space sharing a common
prefix.  Relaxing this constraint leads to a more
general functional, the effective complexity.

Gell-Mann and Lloyd\cite{GL96} describe a procedure
for determining 'Effective Complexity' which extends
the principles of of maximum entropy\cite{Jaynes} to
complexity theory.  The total information functional
$\Sigma$ is defined as the sum of the Shannon
information of an ensemble Z, of which the string X is
a member, and another argument, the effective
complexity, Y, the K-complexity of this ensemble.  By
minimizing total information subject to arbitrary
constraints $f(X)=c$, which incorporate any prior
information known about the system, the most likely
configuration of the ensemble may be determined.
\begin{equation} \Sigma = Y + H(Z) = K(Z) +
H(Z)\end{equation}

This expression minimizes complexity and maximizes
uncertainty. Typically, the total information is
minimized by the Kolmogorov complexity\cite{GL96} and
these quantities are within a few bits of $K = Y +
H$\cite{GM03}.  The relationship between K, Y, and H
may be characterized in terms of input to a computer
program.  The effective complexity Y represents a
fixed deterministic algorithm, and the entropy H
represents the information content of an arbitrary
initial condition chosen as input for that
algorithm\cite{GM03}.  Together, these represent the
minimal total information content needed for the
output of the program.  In the absence of any
additional constraints, this is tyically the
Kolmogorov complexity K.

When macrostates are cylinder sets, which are
coarse-grained partitions of phase space, the
effective complexity becomes identical to the
algorithmic entropy.  In contrast to algorithmic
entropy, effective complexity applies generally to any
macrostates, which need not be compact volumes of
state space and their string representations don't
generally share a common prefix.  Such macrostates may
represent any set of objects equivalent under an
arbitrary relation; however, coarse-grained
macrostates are a very special case which lead to
algorithmic entropy and an alternative definition of
the prefix complexity.  Note that when algorithmic
entropy is generalized to ensembles of arbitrary
measure, it becomes equivalent to the effective
complexity.

Finally, we note that the Kolmogorov complexity is not
generally calculable due to non-halting
programs\cite{CH87}, and, moreover, a binary
computation system is only optimal for representing
powers of two.  Though non-constructive, Kolmogorov
complexity is a useful conceptual device which
simplifies the reasoning of many proofs, e.g.
demonstrating the incompleteness of axiomatic systems
or the limits of inductive reasoning\cite{LV97}.

\section{Macroscopic Equivalence of Microstates}
Macrostates may be described using the simplest
representation of an equivalent microstate.  This
measure has an important application to 'lossy' data
compression, as its objective is to find the simplest
representation which is equivalent to a more complex
datum.  This is a function of the particular observer
or classifier involved, which we may characterize by
an equivalence relation, $P$.

The simplest string capable of reconstructing an
object equivalent to $X$ is the simplest definition of
an object belonging to the equivalence class $X/P$;
since no shorter string appears equivalent to the
observer, this code is optimal.  This establishes a
formal theoretical limit for the performance of the
so-called lossy data compression algorithms prevalent
in digital media.  Beyond this level, information
about microscopic state constitutes irrelevant noise.
Discarding this irrelevant microscopic data boosts the
signal-to-noise ratio perceived by the macroscopic
observer, which facilitates the inference and machine
learning of macroscopic signals. Macroscopic
equivalence relations arise naturally in the lossy
compression of perceptual data - images, audio, and
video - as the objective of such algorithms may be
phrased as a search for shorter representations which
are indistinguishable to an observer represented by
class $P$.

The observer or classifier $P$ groups
indistinguishable objects into equivalence classes,
with a finite but large number of objects falling into
each equivalence class.  Consider the equivalence
relation $P$ on the set of strings as a function which
maps string representations of microstates to
observable macrostates.  $X$ is indistinguishable from
$Y$ if and only if microscopic states $X$ and $Y$ are
congruent modulo the equivalence class $P$.

A string $X$ represents the microstate of an object or
ensemble, and its equivalence class $X/P$ is the
macrostate of the object/ensemble as observed under
$P$.  For the purposes of this paper, set membership
in class $X/P$ is formally presumed to be determinable
by an Oracle for $P$ - a Turing machine may ask the
Oracle a true/false question to determine set
membership in $X/P$ in a single operation.  In
general, $P$ may take any form.  The equivalence
relation may be endowed with arbitrary criteria so
long as these criteria provide consistent
classification.  The canonical example of classical
thermodynamics involves measurement at a particular
scale, resulting in $P$ which partitions the phase
space of the system at a characteristic length scale.
$P$ may specify, for example, a neural net or other
classifier, time scales, statistics from human
observations, or other factors.

\section{The Complexity of an Equivalence Class}

We introduce a new complexity metric for an object X,
the Kolmolgorov complexity of the simplest object
equivalent to X under the relation P(). $S_{f}(X/P)$
is a measure of the descriptive complexity of an
equivalence class of macroscopic objects.
\begin{equation} S_{f}(X/P) \equiv \min_{Y \in
P(X)}{K_{f}(Y)} \end{equation}

We refer to $S_{f}(X/P)$ as the complexity of a
macroscopic state P, the macrostate complexity, or
simply the macrocomplexity.  These are macrostates in
the sense of classical thermodynamics; as such, the
logarithm of their cardinality is the Boltzmann
entropy S.  $S_{f}(X/P)$ is the minimum Kolmolgorov
complexity of any string equivalent to X, the length
of the shortest computer program which terminates with
output in the class P(X).  K(Y), then, could also be
used as a minimal description of the macroscopic
equivalence class P(X).  K(Y) represents the shortest
description macroscopically equivalent to X, which is
the optimal information-losing ('lossy') data
compression of string X (on computer f).

In contrast to the Kolmolgorov Structure Function,
which produces the minimal Sufficient Statistic as a
superset of x given the desired complexity of this
set, the macrocomplexity is a function of x and its
superset P(X).

To simplify the expression, we substitute the
definition of $K_{f}(X)$ into the definition of
S(X/P), which reduces to: \begin{equation} S_{f}(X/P)
\equiv \min_{f(p) \in P(X)}{|p|} \end{equation}

This looks similar to the definition of the Kolmogorov
complexity, but the equality in the argument has been
replaced by an equivalence.  The macrocomplexity
$S_{f}(X/P)$ is a function of the microstate X and
equivalence relation P, in contrast to Kolmogorov's
C-complexity, which depends only on X.  Clearly, $S
\leq C$.  In fact, the difference between the
macrocomplexity $S_{f}(X/P)$ and the C-Complexity of a
typical state is close to Boltzmann's entropy
function.

\section{Boltzmann Entropy and Optimal
Information-Losing Codes}

The entropy of multimedia data is typically high, its
string representations are irregular and nearly
incompressible by universal (lossless) data
compression algorithms\cite{SS00,RY01,RI03,SA97}, so
the output of such algorithms is not significantly
shorter than the original data.  An effective lossy
compression algorithm, on the other hand, minimizes
description length within an equivalence class whose
elements are indistinguishable to a macroscopic
observer or other equivalence class $P$, which may
allow significant savings.

As a concrete example, consider lossy MPEG Level 3
(MP3) audio compression, which typically provides
higher levels of compression of music than the
universal Lempel-Ziv \cite{LZ76,LZ78} compression
algorithm. MP3 frequently compresses music recordings
90\%, whereas Lempel-Ziv's compression ratio of raw
music data is often close to zero.  The reason such an
improvement is possible, given entropic coding limits,
is that the human nervous system discards large
amounts of irrelevant perceptual
data\cite{SA97,KMS06}.  As a result, the classes of
objects which are indistinguishable to humans often
have many members, which, in turn, leads to the
existence of shorter descriptions.  By refining this
notion, we will elucidate the role of Boltzmann
entropy functions in macroscopic observation and lossy
data compression.

For media such as audio or video which mimic the
sensory channels of a macroscopic human observer, the
amount of regularity or redundancy is often low in
comparison to the length of the string.  Not much
compression is possible, so $C_{f}(X) \approx |X|$. In
this case, X is regarded as mostly random or
chaotic\cite{CO91,LV97}.  Regardless of the string X,
the size of the class $X/P$ naturally affects the
existence of simpler equivalent descriptions.  If the
criteria for P are very restrictive, then it may be
that \begin{equation} S_{f}(X/P) \approx C(X)
\end{equation}

That is, any description of a macrostate requires
specification of nearly the entire microstate.  We
will focus our attention on the more interesting case,
when $X/P$ contains simpler microstates equivalent to
a typical element X: \begin{equation} S_{f}(X/P) <
C(X) \end{equation}

This is the case when lossy compression is practical,
for example, with many digital audio and video
recordings. In such cases, the Boltzmann entropy of
$X/P$ is comparable to the difference between the
macrocomplexity and K-complexity.  This may be
demonstrated via the universal probability measure.
Let us define the universal probability of an
equivalence class: \begin{equation} U(X,P) =
\sum_{f(p) \in X/P}{ 2^{-|p|} } \end{equation}

Here the programs p are implied to be self-delimiting
prefix codes, and hence the relation involves
K-complexity rather than C-complexity.  This may be
rewritten as the sum of the individual universal
probabilities $U(X_{i})$ for each string $X_{i}$
belonging to the class $X/P$: \begin{equation} U(X,P)
= \sum_{i=1}^{|X/P|}{U(X_{i})} \end{equation}

To first order, the universal probability of programs
having X as output is dominated by the shortest
program and may be approximated by \begin{equation}
U(X) \approx 2^{-K(X)} \end{equation}

The universal probability of programs congruent to
$X/P$ may be expressed as \begin{equation} U(X,P)
\approx 2^{-S(X/P)} \end{equation}

The relative frequency of programs with output in the
class $X/P$ over programs whose output is X, then, is
the ratio of these two measures.  The universal
probability of microstate X given that X is in $X/P$
becomes \begin{equation} U(X|X/P) =
\frac{U(X)}{U(X,P)} =
\frac{U(X)}{\sum_{u=1}^{|X/P|}{U(X_i)}}
\end{equation}

or, taking the leading terms in each series, we have,
to first order, \begin{equation} U(X|X/P) \approx
2^{S(X/P)-K(X)} \end{equation}

In a classical statistical ensemble, each of the
$|X/P|$ microstates of the system are equally likely,
with probability $\frac{1}{|X/P|}$.  These
probabilities do not directly correspond to the
universal probabilities. The latter are the
probabilities of obtaining a string as the output of a
random program on a certain Turing machine, and the
former are simply the probabilities directly implied
by the length of the string.  Directly equating the
universal probability and the likelihood is not
appropriate.

However, we may characterize an typical element $X$
whose universal probability is close to its mean value
of $\frac{1}{|X/P|}$.  For such a typical element:
\begin{equation} U(X|X/P) = \frac{U(X)}{U(X,P)} =
\frac{U(X)}{\sum_{u=1}^{|X/P|}{U(X_i)}} \approx
\frac{1}{|X/P|} \end{equation}

Substituting, we see that the cardinality of the
macrostate obeys, to first order, \begin{equation}
U(X|X/P) = \frac{1}{|X/P|} \approx 2^{S(X/P)-K(X)}
\end{equation}

After taking logarithms and inverting the sign, we
obtain a relation involving the the Boltzmann entropy
of the macrostate, which is the logarithm of the set
cardinality: \begin{equation} S = \log{|X/P|} \approx
K(X)-S(X/P) \end{equation}

The Boltzmann entropy of the macrostate is seen to be
the difference between the prefix complexity K and the
macrocomplexity $S(X/P)$ for a typical element of
$X/P$ which occurs with approximately average
probability. Entropy represents the additional
information needed to specify a typical microstate, of
complexity $K(X)$, provided the description of its
macrostate having complexity $S(X/P)$.  This holds for
'shallow' objects, where the universal probability is
dominated by the shortest program, but need not be the
case for 'deep' objects which reveal their structure
in a slow convergence to the universal probability.

If the system has multiple macrostates, rather than a
single equilibrium macrostate, then a different
probability measure may apply.  The uniform recursive
probability measure for strings of length $|X|$,
$\mu=2^{-|X|}$, implies: \begin{equation}
\frac{|X/P|}{2^{|X|}} \approx 2^{K(X)-S(X/P)-|X|}
\end{equation}

This measure effectively shifts the discrete Boltzmann
entropy by a normalization constant $|X|$:
\begin{equation} S = \log{\frac{|X/P|}{2^{|X|}}}
\approx K(X)-S(X/P)-|X| \end{equation}

This form is related to universal randomness
tests\cite{GA94,LV97}.  To first order, the sum of the
macrocomplexity and Boltzmann entropy (which is also
the total information $\Sigma$) may be expressed as
the Martin-L\"{o}f universal randomness test
$K(X)-|X|$: \begin{equation} \Sigma = S+S(X/P) \approx
K(X)-|X| \end{equation}

Such relationships are known to relate algorithmic
entropy\cite{LV97} and prefix complexity, which may be
expressed as a special case of macrocomplexity.  The
macrocomplexity provides statistical and thermodynamic
bounds for the optimal performance of lossy data
compression, just as Shannon information limits exact
universal compression.  Furthermore, since the
macrocomplexity is the Kolmogorov complexity of an
equivalent element, it may be used (or approximated)
for minimum description length inference in problems
of pattern recognition and artificial intelligence.

\section{Effective Complexity and Algorithmic Entropy
of Macrostates}

Like many complexity measures\cite{GM03},
macrocomplexity is closely related to effective
complexity and algorithmic entropy\cite{LV97}.  These
measures agree, under appropriate conditions, with
macrocomplexity as the best information-losing data
compression of a string $X$ as judged by a classifier
$P$.

The entropy of an unconstrained, discrete set is the
logarithm or Boltzmann entropy function $S=\log |X/P|$
of the macrostate, so the total information becomes:
\begin{equation} \Sigma = Y + H(X/P) = Y + \log |X/P|
\end{equation}

Where $Y$ is the K-complexity of the macrostate $X/P$.
For the case of a typical element X, the total
information $\Sigma$ is close (within a few
bits\cite{GM03}) to the K-complexity X.
\begin{equation} K(X) \approx Y + \log |X/P|
\end{equation}

This level of effective complexity is typical of the
equivalence class $X/P$.  Hence, for typical elements,
the effective complexity is: \begin{equation} Y
\approx K(X) - \log |X/P|  \end{equation}

This is approximately equal to the first-order
approximation to the macrostate complexity obtained in
the previous section: \begin{equation} S(X/P) \approx
K(X)-\log{|X/P|} \end{equation}

So, $Y \approx S(X/P)$ is the complexity typically
perceived by an observer or some other classifier
described by the macrostate P.  The correspondence may
be seen to hold more generally by considering
macroscopic equivalence in terms of Turing
equivalence.  The S-complexity may be alternately
defined using the complexity of a computer-observer
system.  In this context, the entropy plays the role
of a constant which relates the complexity of programs
on computer f to those of a Turing-equivalent
computer-observer system, g.  Specifically, g applies
to its input program the instructions of computer f
followed by the mapping to the equivalence class P(X),
that is, $g() = P(f())$. This mapping loses
information, so the complexities obey:
\begin{equation} K_{f}(X) = K_{g}(P(X)) + C
\end{equation}

As we have seen, for typical elements, the additive
Turing equivalence constant, $C$, is approximately the
Boltzmann entropy $S$.  It represents the amount of
Kolmolgorov complexity lost by the computer-observer
system, $g=P(f)$, as compared to the standard
computer, $f$.  This allows an alternative definition
of macrocomplexity - the K-complexity of a macroscopic
equivalence class $P(X)$ on the computer-observer
system $g$: \begin{equation} S_{f}(X,P) \equiv K_{g}(P(X))
\end{equation}

The macrostate complexity, originally defined in terms
of a standard computer and an observer, is now the
complexity of the macrostate on a computer system
which incorporates the observer.  This is an effective
complexity, the K-complexity of P(X).  In this way, we
split the total information $K_{f}(X)$ into an
effective complexity $K_{g}(X)$ and a Turing
equivalence constant C, a function of P which plays
the role of the entropy.

Equivalently, macrocomplexity may be regarded as a
generalization of algorithmic entropy.  Endowing the
algorithmic entropy with an arbitrary metric
generalizes the prefix complexity\cite{LV97} to
macrostates which do not necessarily share common
prefixes; in this case, macrocomplexity arises from
the algorithmic entropy of arbitrary sets having
uniform metrics.

\section{Calculating the Complexity of a Macrostate}

The first step in a consideration of macrocomplexity
is to specify the observer or classifier P.  Once this
is done, a calculation or estimate of complexity may
be desired.  Kolmogorov complexities are not generally
calculable\cite{CH87}, so unless the class $X/P$
contains objects with short string representations,
exact calculation of $S_{f}(X/P)$ could be impossible.
Even if one discounts the halting problem and uses the
finitely calculable resource-limited
complexities\cite{LV97}, the number of enumerable
strings to consider is potentially daunting.

Practical approximation of $S_{f}(X/P)$, however, may
be fairly simple, given one or more lossy compression
algorithms offering good performance as judged by the
classifier P.  Estimation of $S_{f}(X/P)$ in this case
amounts to tuning the lossy algorithms to minimize
length without perceptible loss, as determined by the
relation P.  Just as universal data compression
algorithms such as Lempel-Ziv\cite{LZ76} may be used
to estimate the algorithmic prefix complexity, K,
existing lossy data compression algorithms allow a
quick estimate of certain macrostate complexities.
These macrostate complexities may then be used to
construct a mutual information function, universal
probabilities, or other statistics.

Of course, if the cardinality of the macrostate,
$|X/P|$, is known or estimable, then $S_{f}(X/P)$ may
be approximated using results obtained relating the
macrocomplexity to the Boltzmann entropy.  Conversely,
knowledge of $S_{f}(X/P)$ may be used to estimate the
cardinality or Boltzmann entropy of an unknown
macrostate (with a known equivalence relation) whose
cardinality might otherwise be difficult or impossible
to count.

\section{Classification of Macrostates}

Given the ability to calculate or approximate
$S_{f}(X/P)$, the minimal descriptive complexity of an
element of the equivalence class $X/P$, we may use it
for the purpose of classification. We may define a
conditional complexity by directly substituting
macrocomplexity in place of K-complexity:
\begin{equation} S((A|B)/P) = S(AB/P)-S(B/P)
\end{equation}

Effectively, this is $K(A|B)$ modulo P, and it
represents the complexity of differences between A and
B which persist even after passing through the
classifier P.  The combination AB is typically chosen
in a way that preserves locality under P.  For
practical resource-limited estimation, AB should be
constructed such that corresponding structural
elements of the objects A and B are 'close' in some
sense of the resulting representation.

Ideally, a similarity measure $D(x,y)$ would have the
properties of a distance metric: $D(x,y) > 0$, $D(x,y)
= D(y,x)$, and $D(x,y) + D(y,z) < D(x,z)$.  One way to
accomplish this would be to symmetrize by
addition\cite{LV97} to $S((A|B)/P)+S((B|A)/P)$ which
results in the macroscopic complexity's equivalent of
a mutual information function\cite{CO91,LV97} modulo
the relation $P$.  However, this combination is not
necessarily the desired minimal distance function, as
there is generally some redundancy between these two
quantities.  On the other hand, the 'max-distance' $E1
= \max(C(A|B),C(B|A))$ is minimal among all such
distances\cite{LV97}, up to an additive constant.  In
terms of macrocomplexity, this is: $D(A,B) =
\max\{S((A|B)/P),S((B|A)/P)\}$

This is the minimum amount of additional data that
must be specified to transform A into an element of
P(B) or B into an element of P(A), with the optimal
transformation being in one of these two directions.
This quantifies the similarity of any two macroscopic
objects and provides a natural framework for the
classification of macrostates.  In this framework,
classification problems reduce to minimizing a sort of
universal invariant distance from X to the class
$P_{i}$, \begin{equation} D_{P_i}(X) = \min_{Y\in P_i}
D(X,Y) \end{equation}

This is evaluated against all macrostates in $P$ to
identify the closest macrostate, $P_{i}$, to a string
whose equivalence class $X/P$ is undefined or unknown:
\begin{equation} Class(X) = \arg\min_{P_i \in P}
D_{P_i}(X) \end{equation}

For example, the recognition of recorded speech as
particular words might evaluate an unknown audio
sample against recordings indexed in a dictionary.
Such patterns may be specified, as would typically be
the case with a dictionary, or they may be defined
based on proximity in equivalence distance.

\section{Comments and Discussion}

The extraction of meaningful information has always
been a problem in the machine recognition of human
sensory input.  Prior to filtering by neural
perceptual classifiers, such input is mostly random,
incompressible, noisy and chaotic.  The perception of
useful information in such a signal involves removal
of irrelevant noise in order to recognize compressible
and learnable patterns.  Lossy data compression
schemes, by their nature, do this, and, coupled with a
macroscopic equivalence relation, allow the practical
estimation of macrocomplexity in some cases.  As lossy
compression algorithms improve, so will approximations
of macrocomplexity, which will improve the quality of
pattern recognition.

As evidenced by the success of lossy perceptual audio
encoding, psychoacousics has become a relatively
mature science.  When perceptual equivalence under $P$
amounts to indistinguishability to a typical human
ear, these psychoacoustic models provide effective
lossy data compression. The resulting macrocomplexity
may be used to perform auditory inference by proximity
in equivalence distance.

Spoken language is richer in information than text,
but the difficulty of extracting this information has
historically limited its utility in analysis.  For
example, in the case of written human language, the
transcription of audio data to symbolic data obviously
loses large amounts of information about cues such as
inflection, tone, and timing.  One could define
macroscopic perceptual classes based on some semantic
equivalence, e.g. $X/P$ could represent recordings of
a particular word.  Psychoacoustics models, however,
properly describe indistinguishability of sounds to
the ear of a hypothetical listener rather than this
sort of higher-order linguistic processing.

As a trivial example of how macrocomplexity is
relevant to inference, consider the filtering of human
speech recorded in a noisy environment.  If the
original recording is, for example, a 48kHz channel
recorded on an idealized microphone, then most of its
data points are irrelevant to the capture of the human
voice, whose frequency response does not exceed some
maximum frequency threshold, typically around 3kHz.
The speech frequency band constitutes a psychoacoustic
model for an observer, $P$, ignorant of the higher
frequencies.  As such, a crude perceptual coding might
simply perform a Fourier transform and discard all
frequencies in the spectrum above (or below) the
audible threshold.  Entropic compression may then be
used to estimate complexities or information distances
using either the filtered spectrum or it's inverse
Fourier transform, the filtered signal. In addition to
reconstructing the signal using less information, this
improves inference regarding speech, as higher
frequency components are not germane to speech
analysis.  The result is an estimate of
macrocomplexity for the specified $P$.

The lossy compression of images and video is a more
difficult problem than audio, as visual processing by
humans is not so well understood as psychoacoustics,
and because images often represent more information
than audio recordings.  Modern image compression is
largely based on wavelets through the JPEG2000
standard\cite{AT05}, which remains the dominant medium
for transmitted image data.  Recent video codecs such
as h.264 video and MPEG-7 audio have been developed
using more realistic psychological
models\cite{KMS06}\cite{RI03}.  These algorithms offer
more effective perceptual coding, greater compression,
and hence more accurate estimates of macrocomplexity
as compared to their predecessors.  As one might
expect based on the discussion here, such algorithms
have improved utility in indexing and retrieval
applications\cite{KMS06}.

Due to the breadth of other definitions of equivalence
classes, the notion of splitting microscopic
complexity into a macroscopic observer and a
macroscopic observation has many possible
implications.  Macrostate complexity may only be
rigorously defined insofar as the macroscopic
equivalence relation $P$ may be well-posed.  A
rigorous and exact calculation of macrocomplexity,
like the Kolmogorov complexity of its microstate, is
not possible beyond trivial cases.  However, for
observers similar to those modeled by existing lossy
data compression algorithms, the use of such a measure
enables pattern recognition which can exceed the
limits posed by classical information theory.

\providecommand{\bysame}{\leavevmode\hbox
to3em{\hrulefill}\thinspace}
\providecommand{\MR}{\relax\ifhmode\unskip\space\fi MR
} 
\providecommand{\MRhref}[2]{%
  \href{http://www.ams.org/mathscinet-getitem?mr=#1}{#2}
} \providecommand{\href}[2]{#2}

\end{document}